\documentclass[twoside]{article}

\newcommand{\DoMemo}{T}

\newcommand{\memnum}{106}
\newcommand{\whatmem}{\AIPS\ Memo \memnum}
\newcommand{\AIPS}{{$\cal AIPS\/$}}

\newcommand{\memtit}{Making Movies from Radio Astronomical Images with
AIPS}

\newcommand{\TEX}{\hbox{T\hskip-.1667em\lower0.424ex\hbox{E}\hskip-.125em
X}}

\newcommand{\boxit}[3]{\vbox{\hrule height#1\hbox{\vrule width#1\kern#2%
\vbox{\kern#2{#3}\kern#2}\kern#2\vrule width#1}\hrule height#1}}
\title{
   \vskip -35pt
   \if T\DoMemo
      \fbox{{\large\whatmem}} \\
      \fi
   \vskip 28pt
   \memtit\\}
\author{C. C. Cheung, D. C. Homan, J. F. C. Wardle, and D. H. Roberts
\\{\it Brandeis University}\\
{\it ccheung,dhoman,wardle,roberts@brandeis.edu}}

\newcommand{\normalstyle}{\baselineskip 4mm \parskip 2mm \normalsize}

\begin{document}

\pagestyle{myheadings}
\thispagestyle{empty}

\if T\DoMemo
   \newcommand{\Rheading}{\whatmem \hfill \memtit \hfill Page~~}
   \newcommand{\Lheading}{~~Page \hfill \memtit \hfill \whatmem}
\else
   \newcommand{\Rheading}{C. C. Cheung et al.\hfill \memtit \hfill Page~~}
   \newcommand{\Lheading}{~~Page \hfill \memtit \hfill C. C. Cheung et
al.}
   \fi
\markboth{\Lheading}{\Rheading}

\vskip -.5cm
\pretolerance 10000
\listparindent 0cm
\labelsep 0cm

\vskip -30pt
\maketitle
\normalstyle

%\affil{Physics Department, MS-057, Brandeis University, Waltham, MA
%02454}
%\email{ccheung,dhoman,wardle,roberts@brandeis.edu}

\begin{abstract}

We present a detailed recipe for making movies from multi-epoch radio
observations of astronomical sources. Images are interpolated linearly in
time to create a smooth succession of frames so that a continuous movie
can be compiled. Here, we outline the procedure, and draw attention to
specific details necessary for making a successful movie. In particular,
we discuss the issues pertaining specifically to making polarization
movies.  The procedure described here has been implemented into scripts in
NRAO's AIPS package (Brandeis AIPS Movie Maker -- BAMM) that are available
for public use (http://www.astro.brandeis.edu).

\end{abstract}

%\keywords{techniques: image processing}

\section{Introduction}

We have long been interested in presenting sequences of polarization
sensitive VLBI (Very Long Baseline Interferometry) images as ``magnetic
movies'' to show the evolution of the parsec-scale magnetic field in the
jets of blazars. Early VLBI polarization experiments revealed complex
magnetic field structure in these sources with dynamic polarization
properties on short time scales (e.g. Brown, Roberts \& Wardle, 1994) and
intra-day variability has been observed in some sources
(Gabuzda et al. 1989, 2000a,b). Now, instruments such as the
NRAO\footnote{The National Radio Astronomy Observatory is a facility of
the National Science Foundation operated under cooperative agreement by
Associated Universities, Inc.} Very Long Baseline Array (VLBA) and the
Very Large Array (VLA) are routinely used to rapidly monitor changes in
source structure in a variety of radio sources, making movies of these
objects possible.

Here we present our method for making total intensity and/or polarization
movies from multi-epoch radio observations. We focus especially on the
issues specific to polarization sensitive images.  This procedure has been
implemented in a script in NRAO's AIPS package, and we have applied it to
make some ``superluminal magnetic movies'' (section~\ref{application}). 
 
\section{Methodology}

In this section, we discuss our procedures for creating movies from a
series of radio {\it images}\footnote{Throughout our discussion, we refer
to the original AIPS files as {\it images} and to those files interpolated
from the {\it images} using our procedure as {\it frames}.} obtained at
differing epochs. We have implemented this procedure into an AIPS script
in the form of a RUN file named BAMM (Brandeis AIPS Movie Maker; see
section~\ref{application}).  In order to fill in ``gaps'' in the sequence,
we interpolate between adjacent images to create a smooth succession of
{\it frames} to combine into a movie. We favor interpolating linearly
between successive epochs over more sophisticated weighting schemes in
which non-adjacent images are factored into the interpolation since linear
interpolation is simple and well defined. Also, we caution against
applying this method to data undersampled in time as one may be misled by
the results (e.g.  ``strobing'' effects). 

\subsection{Making the Movie Frames}

The data assumed here are AIPS image files derived from UV data using the
same cell sizes and field of view. Calibrated $I$, $Q$, and $U$
images\footnote{$I$, $Q$, and $U$ correspond to the Stokes parameters for
total intensity and linear polarization; $Q$ and $U$ images are only
necessary for polarization movies.} are read into NRAO's AIPS reduction
package and the following procedures are followed. 

\subsubsection{Total Intensity Movies}
\label{imovie}

\begin{enumerate}

\item 
\label{stepi1}
All the images need to be restored to a common beam size. We restore
the CLEAN components of each image onto a blank image with the task
RSTOR. In order to avoid false interpretation of moving features due to
effects of changing resolution, we use the largest beam amongst the images
for all frames.

\item 
\label{stepi2} 
Images are then aligned relative to a common stationary (total intensity)
feature, and a common image size (field of view) is chosen. The stationary
feature can be, for instance, the (presumed) stationary core of a quasar
jet or a bright feature in a maser source. We use the task LGEOM to
perform this step. 

\item
\label{stepi3}
Restored and aligned images are linearly interpolated with the task
COMB to output a user specified number of frames at evenly spaced
epochs. The number of frames determines the time elapse between the
resultant movie frames and the weighting given to adjacent images
in the interpolation.

\end{enumerate}

\noindent A total intensity movie can now be compiled from these frames
(section~\ref{compile}). 

\subsubsection{Polarization Movies}

Here, we discuss additional steps necessary for making polarization
movies. Proceed with steps~\ref{stepi1} \&~\ref{stepi2} from
section~\ref{imovie}, this time using all three Stokes ($I$, $Q$, and $U$)
images. Stokes $V$ (circular polarization) images can be treated
identically to $I$ images as described in section~\ref{imovie}. After
aligning the $I$ images, the corresponding $Q$ and $U$ (and $V$) images
must be shifted by the same amount.

\begin{enumerate}

\item 
If rotations to the tick-marks that illustrate the orientation of the
electric field vectors need to be made\footnote{If not, proceed to
item~\ref{laststep} using $Q'=Q$ and $U'=U$.} (e.g. accounting for Faraday
rotation by adding a correction angle, $\theta$, or changing the
tick-marks to show the orientation of the magnetic field
vectors\footnote{Assuming negligible Faraday rotation,
$\theta=\frac{\pi}{2}$.}), the total polarized intensity
($p=\sqrt{Q^{2}+U^{2}}$)  and electric field vector orientation
($\chi'=\frac{1}{2}\arctan{\frac{U}{Q}} + \theta$) images at each epoch
are made first from the restored and aligned $Q$ and $U$ images with the
task COMB. This is done in order to avoid the problem of having tick-marks
rotating improperly later. 

\item
The resultant $p$ and $\chi'$ images are used to make new $Q$ and $U$
images\footnote{While $Q$ and $U$ have changed in order to rotate the
tick-marks, $p$ ($=\sqrt{Q'^{2}+U'^{2}}$) remains unchanged.}
(eqns~\ref{eqn:q} \&~\ref{eqn:u})  using the task COMB, via:

\begin{equation}
\label{eqn:q}
Q'=p*\cos{2\chi'},
\end{equation}

\begin{equation}
\label{eqn:u}
U'=p*\sin{2\chi'}.
\end{equation}

\item
\label{laststep}
Lastly, the $I$, $Q'$, and $U'$ images are linearly interpolated
separately with the task COMB. These images are then used to make the $I$,
$p$, and fractional polarization ($m=p/I$)  movie frames.  The length of
the tick-marks on the $p$ frames can be made uniform or proportional to
the magnitude of $m$ or $p$.

\end{enumerate}

\noindent 
Our AIPS script, BAMM, will implement all of these steps
(section~\ref{application}) after the user specifies the epochs of
observations and the positions of a stationary feature. The individual
movie frames are ready for inspection with standard AIPS tools, or
the procedure may be repeated with different parameters if the user is not
satisfied with the output.

\subsection{Compiling the Movies}
\label{compile}

At this point, the user is satisfied with the quality and duration of the
movie frames and wishes to export them into a movie outside of AIPS.  The
user may proceed in a number of ways. For contour movies, frames can be
written out of AIPS in postscript format. It is important that identical
contour levels (in mJy/beam) be used for all the frames. The files are
then converted to GIF format which can be done with software such as XV,
Adobe Photoshop, or ps2gif.

For color movies, the frames can be displayed on the AIPS TV and the
images are captured with the task TVCPS or even XV. Also, the frames can
be exported to software packages like the Caltech VLBI package or IDL. The
same display scale should be used for all the frames. Scripts to implement
these steps in AIPS are available upon request.  We combine the movie
frames (now in a standard graphics file format) into movie clips using
Adobe Premiere for export to video tape, animated GIFs, or Quicktime
movies. Other formats are supported by Premiere. Also, IDL has been used
to compile movies similar to ours (B.G. Piner, personal communication).

\section{Application: VLBA Polarization Monitoring of 12 Blazars}
\label{application}

We have applied our procedure to images of twelve blazars monitored at 15
and 22 GHz with the VLBA in 1996 by the Brandeis Radio Astronomy Group,
yielding six epochs of polarization sensitive images separated by about
two months each. Details of this monitoring program are discussed by Homan
et al.  (2001) and references therein. These data are sufficiently well
sampled in time that a simple linear interpolation between two adjacent
epochs yielded satisfactory results. The movies show clear superluminal
motion, and reveal the complexity of the evolution of the magnetic field
structure in these sources. These movies have provided a highly visual
medium in which to show our work for scientific talks, general lectures,
and are on our WWW site.\footnote{http://www.astro.brandeis.edu} BAMM,
instructions, other AIPS macros, and our ``Superluminal Magnetic Movies''
are available to the public by following the links in our WWW site or by
contacting one of the authors (CCC).

\section{Acknowledgments}

We are grateful to Jody Attridge and Roopesh Ojha for useful discussions. 
This work was supported by NSF grants AST 95-29228 and AST 98-02708. CCC
received grants from the following Brandeis programs: Ronald E. McNair
Scholars Program, Jerome A. Schiff Undergraduate Fellows Program, Louis,
Frances and Jeffrey Sachar Fund, and the Undergraduate Research Program. 

\section{References}

Brown, L.F., Roberts, D.H., \& Wardle, J.F.C. 1994, ApJ, 437, 108

Gabuzda, D.C., Wardle, J.F.C., \& Roberts, D.H. 1989, ApJ, 338, 743

Gabuzda, D.C., Kochenov, P.Y., Cawthorne, T.V., \& Kollgaard, R.I. 2000a,
MNRAS, 313, 627

Gabuzda, D.C., Kochenov, P.Y., Kollgaard, R.I., \& Cawthorne, T.V. 2000b,
MNRAS, 315, 229

Homan, D.C., Ojha, R., Wardle, J.F.C., Roberts, D.H., Aller, M.F., Aller,
H.D., \& Hughes, P.A. 2001, ApJ, 549, 840

\end{document}